\documentstyle[graphicx]{JHEP}

\newcommand{\real}{{{\rm I} \kern -.19em {\rm R}}}

\title{The Graceful Exit in  Pre-Big Bang String Cosmology}
\author{C. Cartier \\ Centre for Theoretical Physics, University of Sussex,
\\ Falmer, Brighton BN1 9QJ, U.~K.
\\ \email{E-mail: c.cartier@sussex.ac.uk}}
\author{ E.J. Copeland \\ Centre for Theoretical Physics, University of Sussex,
\\ Falmer, Brighton BN1 9QJ, U.~K.
\\ \email{E-mail: e.j.copeland@sussex.ac.uk}}
\author{R. Madden \\ IHES, Le Bois-Marie, Route de Chartres 35,
\\ 91440 Bures-sur-Yvette, France.
\\ \email{E-mail: madden@ihes.fr}}

\abstract{
We re-examine the graceful exit problem in the pre-Big Bang scenario of string
cosmology, by considering the most general time-dependent classical correction
to the Lagrangian with up to four derivatives. By including possible forms for
quantum loop corrections we examine the allowed region of parameter space for
the coupling constants which enable our solutions to link smoothly the two
asymptotic low-energy branches of the pre-Big Bang scenario, and observe that
these solutions can satisfy recently proposed entropic bounds on viable
singularity free cosmologies.}


\preprint{SUSX-TH-99018 \\ hep-th/9910169}

\begin{document}
\unitlength=1mm

\section{Introduction}
It is generally accepted that standard cosmology provides a
consistent picture of the evolution of the Universe  from the
period of primordial nucleosynthesis to the present. However, as
we extrapolate further into the past, our knowledge becomes less
certain as we appear to be inevitably led to an initial curvature
singularity \cite{Hawking1}. The emergence of string theory as the
favoured candidate to unify the forces of nature has led a number
of authors to investigate the cosmology associated with it (for a
recent review see \cite{ejc}). One such approach has been
pioneered by Veneziano and his collaborators, and employs the rich
duality properties present in string theory
\cite{Brandenberger1,Veneziano1,Tseytlin1,Veneziano3}, and leads
to a class of solutions which effectively talk about a period
before the Big Bang, the pre-Big Bang scenario
\cite{Veneziano3,Veneziano2}. The Universe expands from a weak
coupling, low curvature regime in the infinite past, enters a
period of inflation driven by the kinetic energy associated with
the massless fields present, before approaching the strong
coupling regime as the string scale is reached. There is then a
branch change to a new class of solutions, corresponding to a post
Big Bang decelerating Friedman-Robertson-Walker era. In such a
scenario, the Universe appears to emerge because of the
gravitational instability of the generic string vacua
\cite{Veneziano4,Gasperini1}. In many ways this is a very
appealing picture, the weak coupling, low curvature regime is a
natural starting point to use the low energy string effective
action. However, there are a number of problems facing the
scenario. One is that of initial conditions, why should such a
large Universe be a natural initial state to emerge from and does
it possess enough inflation before entering the strong coupling
regime \cite{Weinberg,Kaloper,Veneziano4}?  The second is how can
the dilaton field be stabilised in the post Big Bang phase? It
must be decoupled from the expansion since variations in this
scalar field correspond to changes in masses and coupling
constants, which are strongly constrained by observation
\cite{Damour1,Campbell1}. A number of attempts have been made to
do this; by including dilaton self-interaction potentials and
trapping the dilaton in a potential minimum\cite{Barreiro} and by
taking into account the back-reaction on the dilaton field from
quantum particle production \cite{Brustein0}. The third is the
graceful exit problem. How can the curvature singularity
associated with the strong coupling regime be avoided, so as to
allow a smooth branch change between the pre-Big Bang inflationary
solution and a decelerating post Big Bang FRW solution? The
simplest version of the evolution of the Universe in the pre-Big
Bang scenario inevitably leads to a period characterised by an
unbounded curvature. No-go theorems prevent the inclusion of a
single potential to catalyse a graceful exit in vacuum-dilaton
cosmology \cite{Veneziano5,Kaloper5,Kaloper6}, although Ellis et
al. recently claimed that non-singular evolutions can be obtained
through the inclusion of a single scalar potential providing the use of an
exotic equation of state \cite{Ellis}. The current philosophy is
to include higher-order corrections to the string effective
action. These include both classical finite size effects of the
strings, and quantum string loop corrections and have already met
with some success \cite{Veneziano7,Brustein0,Foffa,divers}. The
effect of backreaction arising from long wavelength modes, where
string theory should be adequately described by general relativity
with a minimally coupled scalar field, confirms that the
qualitative effect of these corrections is compatible with an
evolution leading towards exit \cite{Ghosh}.

The motivation behind this paper is to investigate the graceful exit issue
by studying in detail a physically motivated action at both the
classical and quantum level which can incorporate
a number of appealing features, such as being able to maintain
scale factor duality (SFD), even
when these higher order corrections are included.

The classical corrections to the low energy effective string action usually 
have associated with them fixed points where the Hubble parameter in the
string frame is constant and
the dilaton is growing \cite{Veneziano7}.
We will see that although physically very appealing,
the SFD invariant action does not drive the evolution of the Universe into a
good fixed point (in agreement with Brustein and Madden
\cite{Brustein2}), and moreover the inclusion of
quantum corrections do not lead to
a smooth exit into the decelerating FRW branch, rather they lead to a regime of
instability.
Fortunately, relaxing the SFD condition leads to many interesting features:
the loop corrections introduce an upper (lower) bound for the curvature in the
String (Einstein) frame, suggesting that a graceful exit is viable in this
context, and indeed we present a number of successful exits.

The paper is organised as follows. In Section 2, we discuss the effective
low energy action of the heterotic string including O($\alpha'$)
corrections arising from the finite size effects. In Section 3 we review
previous studies of graceful exit using a truncated form of these corrections
and then make a comparison with the full classical correction.
A detailed analysis is presented showing the region of
parameter space which admit classical fixed point solutions. We extend
the analysis by including possible one- and two-loop
quantum corrections. As expected these turn out to be important
in order to obtain a
successful graceful exit to the pre-Big Bang scenario
\cite{Brustein0,Brustein2}.
Particle creation is then used to stabilise the dilaton in the post-Big Bang
era. Finally in Section 4 we summarise our main results.

\section{String effective action}

The pre-Big Bang scenario is an inflationary model starting with a generic state
of extremely weak coupling and curvature, pictured as a gravitational collapse
in the Einstein frame. This rather trivial asymptotic past state is followed by
a dilaton-driven kinetic inflation phase which has to be long enough to solve
the different cosmological problems inherent to standard cosmologies
\cite{Veneziano4}. Later on, this superinflationary period should be smoothly
connected to the FRW regime, characterised by a decelerating expansion and a
frozen or slowly evolving dilaton, whose present expectation value gives rise to
the universal
gravitational constant.

We shall take as our starting point the minimal $4-$dimensional string effective
action:
\begin{eqnarray}
\Gamma^{(0)} &=& \frac{1}{\alpha'}\int d^{4}x \sqrt{-g} {\cal L}^{(0)} \nonumber
\\
 &=& \frac{1}{\alpha'}\int d^{4}x \sqrt{-g} e^{-2 \phi} \Bigl\{ R +
 4 (\partial_{\mu} \phi)^2 \Bigr\},
\label{Low}
\end{eqnarray}
where we adopt the convention $(-,+,+,+)$,
$R^{\mu}_{\;\nu\lambda\rho}=\partial_{\rho}\Gamma^{\mu}_{\nu\lambda}+\dots$ and
$R_{\nu\rho} = R^{\lambda}_{\;\nu\lambda\rho}$ and set our units such that
$\hbar = c = 16 \pi G = 1$. By low-energy tree-level effective action, we mean
that the string is propagating in a background of small curvature and the
fields are weakly coupled. However, the evolution from the pre-Big Bang era to
the present is understood to be characterised by a regime of growing couplings
and curvature. This means that the Universe will have to evolve through a phase
when the field equations of this effective action are no longer valid. Hence,
the low-energy dynamical description has to be supplemented by corrections in
order to reliably describe the transition regime.

The finite size of the string will have an impact on the
evolution of the scale factor when the curvature of the Universe reaches a
critical level, corresponding to the string length scale
$\lambda_{S} \sim \sqrt{\alpha'}$ (fixed in the string frame), and such
corrections
are
expected to stabilise the growth of the curvature into a de-Sitter like  regime
of constant curvature and linearly growing dilaton \cite{Veneziano7,Brustein0}.
Eventually the dilaton will play a major role, and since the loop expansion is
governed by powers of the string coupling parameter $g_{S} = e^{\phi}$,
these quantum corrections will modify dramatically the evolution when we reach
the strong coupling region \cite{Brustein0,Brustein2}. This should
correspond to the stage when the Universe completes a smooth transition to the
post-Big Bang branch, characterised by a fixed value of the dilaton and a
decelerating FRW expansion. One of the unresolved issues of the transition
concerns whether or not the actual exit takes place at large coupling,
$e^{\phi} \ge 1$. If it occurred whilst the coupling was still
small,
then we would be happy to use the perturbative corrections we are adopting.
However, if the Universe is driven into the strong coupling regime before the
exit proceeds then we might expect to have to adopt a
different approach which involves the use
of non-perturbative string phenomena. Such a possibility has recently been
proposed in the context of M-theory \cite{Foffa,Maggiore2}.

The type of corrections we will be considering involve truncations of the
classical action at order $\alpha'$. Although a field redefinition mixing the
different
orders in $\alpha'$ does not change the physics if one considers all orders in
$\alpha'$, it inevitably leads to amBiguities when some orders are truncated.
This means that the cosmological evolution arising from such
actions truncated at
order $\alpha'$ should really only be considered as an indication of the
possible
cosmological behaviour.
The most general
form for a correction to the string action up to fourth-order in
derivatives has been presented in refs
\cite{Meissner,Kaloper2}:
\begin{eqnarray}
\Gamma^{(C)} &=& \frac{1}{\alpha'}\int d^{4}x \sqrt{-g} {\cal L}^{(C)} \nonumber
\\
&=& k\lambda_{0} \int d^{4}x \sqrt{-g} e^{-2\phi} \Bigl\{ a R^{2}_{GB} + b
\Box \phi (\partial_{\mu} \phi)^{2} \nonumber \\
&&\hspace{3.6cm} + c \Bigl\{ R^{\mu\nu}-\frac{1}{2}g^{\mu\nu}R\Bigr\}
\partial_{\mu} \phi \partial_{\nu} \phi + d (\partial_{\mu} \phi)^{4} \Bigr\},
\label{Class}
\end{eqnarray}
where the parameter $\lambda_{0}$ allows us to move between different string
theories and we will set $k \lambda_{0} =-\frac{1}{4}$ to
agree with previous studies of the Heterotic string
\cite{Veneziano7}. $R^{2}_{GB} = R_{\mu\nu\lambda\rho}
R^{\mu\nu\lambda\rho} -4 R_{\mu\nu}R^{\mu\nu} + R^2$ is the Gauss-Bonnet
combination which guarantees the absence of higher derivatives. In fixing the
different parameters in this action we require that it reproduces the
usual string scattering amplitudes \cite{Metsaev}.
This constrains the coefficient of $R_{\mu\nu\lambda\rho}^2$ with the
result that the pre-factor for the Gauss-Bonnet term has to be $a=-1$.
But the Lagrangian can still be shifted by field redefinitions which
preserve the on-shell amplitudes, leaving the three remaining coefficients
of the classical correction
satisfying the constraint
\begin{equation} \label{constraint}
2 ( b+c) + d = - 16 a.
\end{equation}

There is as yet no definitive calculation of the full loop expansion of string
theory. This is of course a Big problem if we want to try and include quantum
effects in analysing the graceful exit issue. The best we can do, is to
propose plausible terms that we hope are representative of the actual terms that
will eventually make up the loop corrections. We believe that the string
coupling $g_{S}$ actually controls the importance of string-loop
corrections, so as a first approximation to the loop corrections we multiply
each term of the classical correction by a suitable power of the
string coupling \cite{Brustein0}. When loop corrections are included, we then
have an effective Lagrangian given by
\begin{equation}
{\cal L} = {\cal L}^{(0)} + {\cal L}^{(C)} + A e^{2\phi} {\cal L}^{(C)} +
B e^{4\phi} {\cal L}^{(C)},
\label{efflag}
\end{equation}
where ${\cal L}^{(0)}$ is given in Eq.~(\ref{Low}) and
${\cal L}^{(C)}$ given in Eq.~(\ref{Class}). The constant
parameters $A$ and $B$ actually control the onset of the loop corrections.

\section{Numerical solutions}
In this section, we consider the impact of both the classical and
quantum corrections of
Eq.~(\ref{efflag}). Naively, we would expect that the latter should only become
significant as we enter the strong coupling regime.
Depending on the value of the dilaton field, the loop corrections
are indeed negligible in the weak coupling regime as the dilaton $\phi
\rightarrow -\infty$, so we expect that
the solutions to the extended equations of motion should initially
be similar to the lowest-order description, with the classical and quantum
corrections introducing an upper bound for the solutions,
thereby regulating their singular behaviour.

Following Brustein and Madden \cite{Brustein0}, we define the parameters $H_S$
as the Hubble expansion in the string frame, $H_E$ as the Hubble expansion in
the Einstein frame and $\dot{\phi}$ the derivative of the dilaton field with
respect to cosmic time, $t_S$. Hence, the solutions to the equations of motion
resulting from Eq.~(\ref{Low}) can be expressed as $2\dot{\phi} =
(3 \pm \sqrt{3})H_{S}$, where the upper (lower) sign refers
to the pre- (post-) Big Bang solution.
Although starting in the perturbative regime, the $(+)$ branch ($H_{S} > 0$,
$\dot{\phi} > 0$ and $H_{E}<0$) evolves toward a curvature singularity,
and the low-energy effective description breaks down.
More precisely, we expect modifications to become significant when the curvature
is of order the Planck length $\lambda_{S}$ and the low-energy effective action 
has to be replaced by one which includes higher-order effects in $\alpha'$.
It has been known for some time that such classical corrections allow a branch 
change 
$(+) \rightarrow (-)$,  which corresponds to a change of sign of the shifted 
dilaton, $\dot{\bar{\phi}} \equiv \dot{\phi} - 3 H_{S}/2$ \cite{Brustein1}. 
Also, to allow a smooth connection to the usual decelerated Friedmann Universe 
where the dilaton will become fixed, the Hubble rate in the Einstein frame has 
to become positive after the branch change. Using the conformal transformation 
relating the String frame to the Einstein frame, 
$g^{E}_{\mu\nu}=e^{-2\phi}g^{S}_{\mu\nu}$, the Hubble expansion in the E-frame 
can be expressed as a function of S-frame quantities,
$H_{E} = e^{\phi} \{H_{S}-\dot{\phi} \}$. This relation allows us to 
define the Einstein
bounce $E_{B} = H_{S} - \dot{\phi}=0$. 
A necessary condition to obtain a successful exit is the violation of the 
null energy condition (NEC) in the Einstein frame, which is associated with
the cross over of the Einstein bounce\cite{Brustein1}.
Thus we see that a number of different conditions must be satisfied for a
successful exit to be obtained. In the figures below, all these constraints
are presented as lines on the plots, i.e. the $(+)$ and $(-)$ branches of the
low energy string action, the branch change and the Einstein bounce.

\subsection{Classical correction}
\begin{figure}[ht]
\begin{center}
\includegraphics[width=6.5cm]{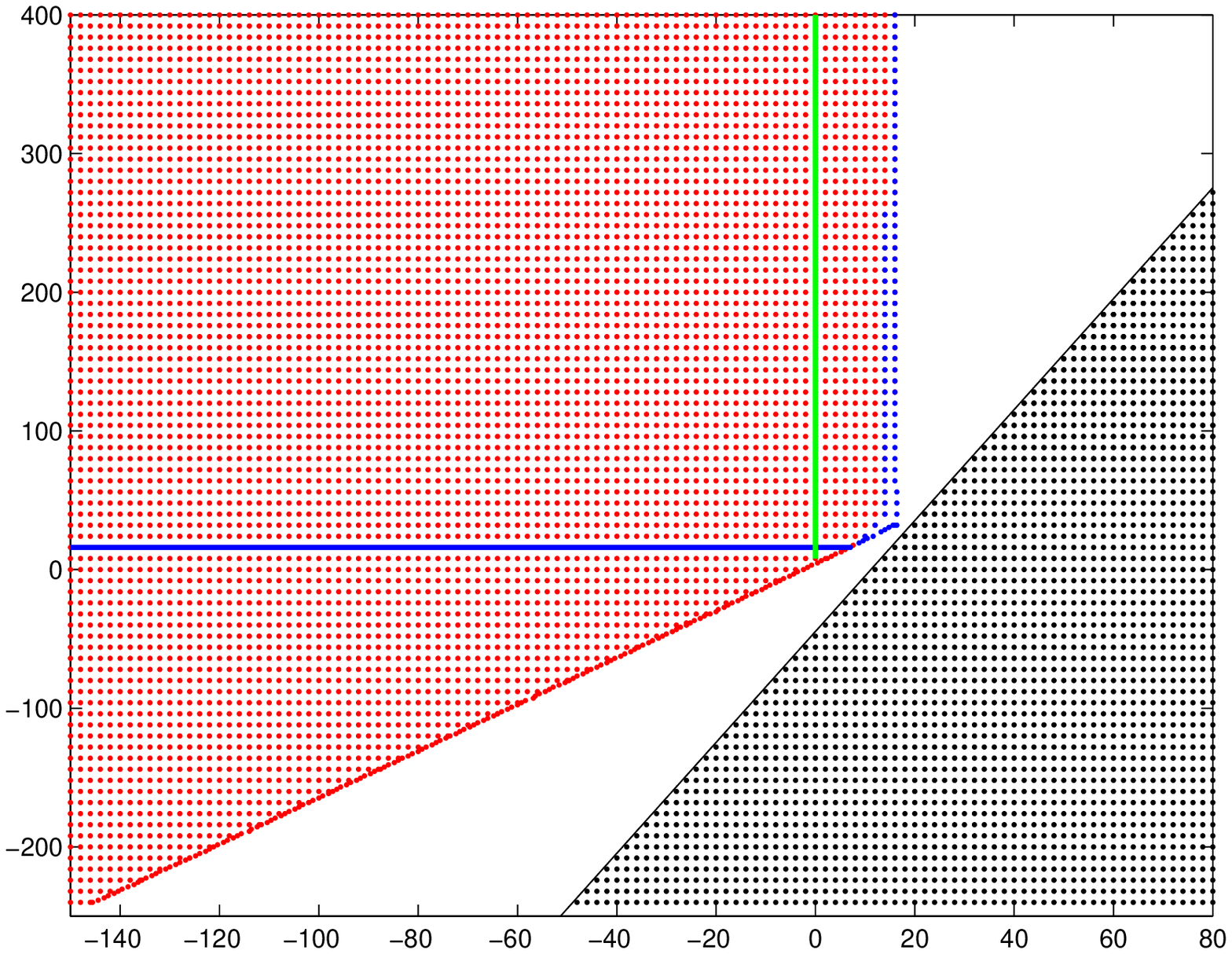}
\hspace{1cm}
\includegraphics[width=6.5cm]{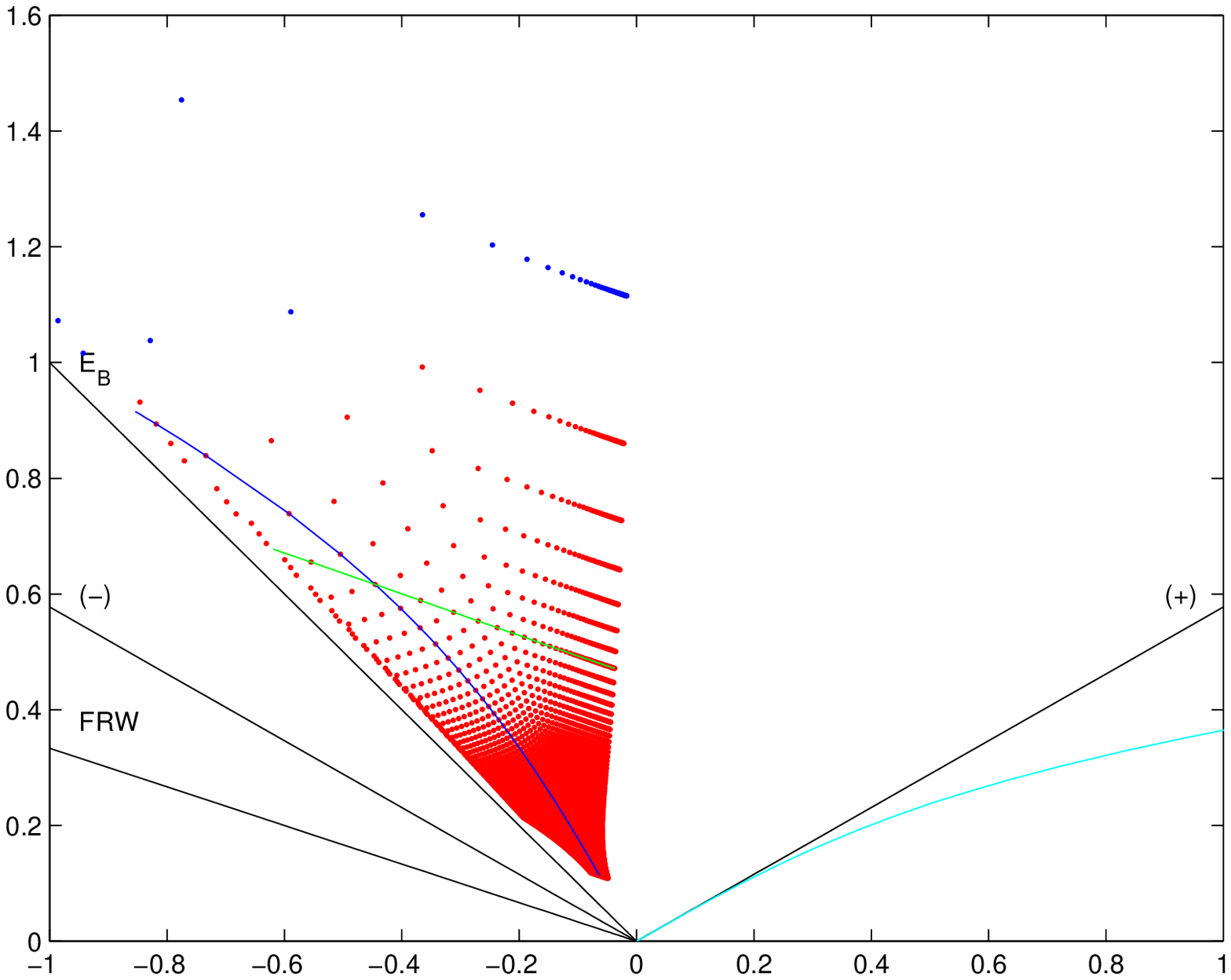}
\caption{On the left, we have represented in the $(c, d)$ plane
the asymptotic state of the evolution when the classical corrections
are included, with $a=-1$ and the remaining parameter $b$ is constrained
by Eq.~(\ref{constraint}). The red dots show the combinations of these
parameters leading to a fixed point, with saturated Hubble parameter
$H={\rm const}$ and a linearly growing dilaton. The black points indicate the
set of coefficients forcing the evolution to head away from the
$\dot{\bar{\phi}} \leq 0$ region. The black line represents the bound
given in \cite{Brustein2}. The figure on the right shows the fixed point
distribution in the plane $(2\dot{\bar{\phi}}, H)$. The green line corresponds
to a constant $c = 0$, whereas the red line stands for $d=16$. The cyan line
shows the evolution for the SFD case with $c=4$ and $d=16$: the curve heads
away from the $\dot{\bar{\phi}} \leq 0$ region, typical of the black dots area 
in
the figure on the left.}
\label{fig0}
\end{center}
\end{figure}

We first recall some aspects of the background evolution when
we restrict ourselves to the classical corrections. Choosing
initial conditions on the pre-Big Bang branch, the equations
of motion derived from Eq.~(\ref{Low}) and Eq.~(\ref{Class}) lead in
general to a regime of constant Hubble parameter $H$ and a linearly
growing dilaton $\dot{\phi} > 0$. This is quantified in figure~\ref{fig0},
where we show the different types of solutions that can be found in the
$(c,d)$ plane, when we choose $a=-1$ and determine $b$ from the
constraint Eq.~(\ref{constraint}).
The distribution of red dots corresponds to values of $(c,d)$ which lead to
standard fixed point solutions, $H={\rm const},~\dot{\phi} > 0$.
The set of blue points (which starts approximatively for $c \geq 12$)
also shows coefficients leading to a fixed point, but with a
Hubble parameter, $H \geq 1$. In particular we see that these points are
bordered by the lines $d \geq  1.68 c +3.62$ and $c \leq 152/9$,
indicating that these represent the constraints on $c$ and $d$
which have to be satisfied in order to obtain satisfactory fixed
point solutions. It is clear that there exists a large region of
parameter space where such solutions are to be found.
Representing these fixed points in the $(2\dot{\bar{\phi}},H)$ plane emphasizes
that for a constant $c=0$ the fixed point moves to smaller values of
$|\dot{\bar{\phi}}|$ and $H$ when we increase the coefficient $d$ (green curve),
whereas $|\dot{\bar{\phi}}|$ and $H$ both become larger when we increase $c$ 
keeping constant $d=16$ (blue curve). This latter curve corresponds to a segment 
of the more general case $b=-c$ where the asymptotic state in the high curvature 
regime is given by the implicit equation:
\begin{eqnarray}\label{fixed}
0 &=& -\frac{3}{16} H (2 \dot{\bar{\phi}}+3 H)  \Bigl\{ 63 \dot{\bar{\phi}} H^5
+ 126 \dot{\bar{\phi}}^2 H^4 - 18 H^4 + 8 \dot{\bar{\phi}}^3 H^3
- 96 \dot{\bar{\phi}} H^3   \\
  &&\hspace{3,4cm}- 176 \dot{\bar{\phi}}^4 H^2 - 144 \dot{\bar{\phi}}^2 H^2
- 144 H \dot{\bar{\phi}}^5 -128 \dot{\bar{\phi}}^3 H - 32 \dot{\bar{\phi}}^6
- 32 \dot{\bar{\phi}}^4 \Bigr\}. \nonumber
\end{eqnarray}
However, we will see later in the analysis that not all values of this curve of 
fixed points can be reached if the initial conditions are chosen on the pre-Big 
Bang branch.

The region to the right of the black line $d \leq  4 c - 44.62$
in figure~\ref{fig0}, represents the range of coefficients which lead to an
evolution heading away from the $\dot{\bar{\phi}} \leq 0$ region. The line was
first proposed by \cite{Brustein2}.
Finally, the intermediate blank area represents the combination of
parameters driving the evolution into a regime of instability for the
scale factor, where $\ddot{a} \rightarrow \infty$. The emergence of
such a region requires further investigation.

We now go on to look at some particular examples including quantum loop
corrections.

\subsection{Minimal case}
\begin{figure}[ht]
\begin{center}
\includegraphics[width=6.5cm]{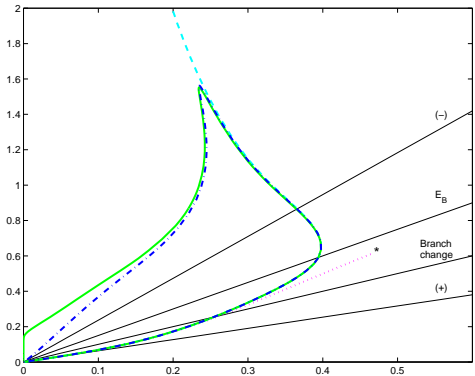}
\caption{Hubble expansion in the S-frame as a function of the dilaton for the 
case $a=-1$, $b=c=0$ and $d=16$. The y-axis corresponds to $H$, and the x-axis 
to $2\dot{\phi}/3$. The initial conditions for the simulations have been set 
with respect to the lowest-order analytical solutions
at $t_{S} = -1000$. The straight black lines describe the bounds quoted in
Section II. The dotted magenta line shows the impact of the classical correction
due to the finite size of the string. A $*$ denotes the fixed point.
The contribution of the one-loop
expansion is traced with a dashed cyan line ($A=4$). The dash-dotted blue line
represents the incorporation of the two-loop correction without the
Gauss-Bonnet combination ($B=-0.1$).
Finally, the green plain line introduces radiation with $\Gamma_{\phi}=0.08$
and stabilises the dilaton.}
\label{fig1}
\end{center}
\end{figure}

The natural setting $b=c=0$ leads to the well-known form which has given rise to
most of the studies on corrections to the low-energy picture. In references
\cite{Veneziano7,Brustein0}, the authors demonstrated that this minimal
classical correction regularises the singular behaviour of the low-energy
pre-Big Bang scenario. It drives the evolution to a fixed point of bounded
curvature with a linearly growing dilaton (the star in figure~\ref{fig1} --
which agrees with the results of \cite{Veneziano7,Brustein0}),
suggesting that quantum loop
corrections -known to allow a violation of the null energy condition
$(p+\rho<0)$- would permit the crossing of the Einstein bounce 
to the FRW decelerated expansion in
the post-Big Bang era. Indeed, the addition of loop corrections
leads to a $(-)$ FRW-branch as pictured in figure~\ref{fig1}. However, we still
have to freeze the growth of the dilaton. Following
\cite{Brustein0}, we introduce by hand a particle creation term of the form
$\Gamma_{\phi} \dot{\phi}$, where $\Gamma_{\phi}$ is the decay width of the
$\phi$ particle, in the equation of motion of the dilaton field and
then coupling it to a fluid with the equation of state of radiation in
such a way as to preserve overall conservation.
This allows us to stabilise the dilaton in the post-Big Bang
era with a decreasing Hubble rate,
similar to the usual radiation dominated FRW cosmology.

\begin{figure}[ht]
\begin{center}
\includegraphics[width=6.5cm]{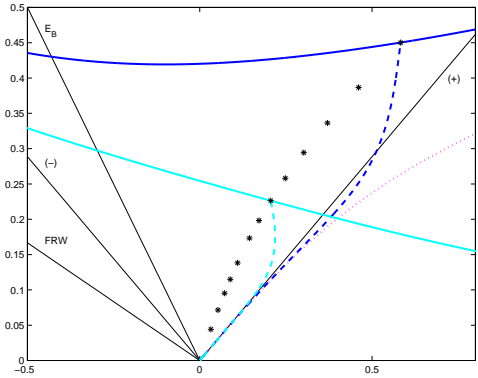}
\caption{Hubble expansion in the S-frame as a function of the shifted dilaton
$(2\dot{\bar{\phi}})$ for the SFD case, with $a=-1$, $b=-c=-16$ and
$d=16$. Straight lines are identical to those in figure~\ref{fig1}. The 
classical
correction makes the curve turn the wrong way (dotted magenta). The inclusion of
the one-loop  correction (dashed) leads to a regime of instability: the
evolution meets the curve corresponding to $\ddot{a} \rightarrow \infty$
(plain). Two cases are explicitly shown: $A=0.5$ in blue and $A=16$ in cyan,
whereas the meeting points for other choices of the pre-factor $A$ are pictured
with stars (*).}
\label{fig2}
\end{center}
\end{figure}

\subsection{Scale factor duality case}
In \cite{Meissner,Kaloper2}, the authors considered the particular
combination $b=-c=-16$ which has the remarkable property of
introducing scale factor duality at the order $\alpha'$
in the correction to the low-energy action at the price of discarding
non-SFD invariant terms of higher order and making a modification to
the definition of SFD at order $\alpha'$.
Unfortunately, as can  be seen in
figure~\ref{fig2}, this choice of parameters does not lead to a successful
exit. The classical correction forces the curve to head away from
the branch changing and exit region.
In fact even including loop corrections it is impossible to reach the branch 
change, given in figure~\ref{fig2} by the line $\dot{\bar{\phi}} = 0$.
Such an observation was previously also made in \cite{Brustein2}. What appears
to be happening is that including the one-loop correction drives the system into
a regime where the acceleration of the scale factor $(\ddot{a})$ diverges. This
is a singular point in the equation of motion and is indicated by the star in
figure~\ref{fig2}.
As discussed in \cite{Brustein2}, it may be that working with higher orders
in the corrections requires further alterations of the form of SFD.

\subsection{The general case}
\begin{figure}[ht]
\begin{center}
\includegraphics[width=6.5cm]{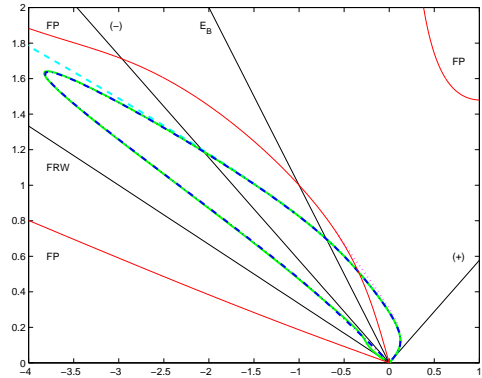}
\caption{Hubble expansion in the S-frame as a function of the shifted dilaton
$(2\dot{\bar{\phi}})$ for the case $a=-1$, $b=-c=-4$ and $d=16$,
with $A=+4$ (dashed cyan) and  $B=-0.3$ (dash-dotted blue). The plain
green line includes the effect of the particle creation, with
$\Gamma_{\phi}=0.2$. The FRW and FP lines represent solutions for
the fixed points given in Eq.~(\ref{fixed}).}
\label{fig3}
\end{center}
\end{figure}

Relaxing the SFD constraint allows us to investigate the full classical and
quantum
correction up to four derivatives and leads to many interesting situations. When
$a=-1$ and $d=16$, this implies that the coefficient of the
curvature-dilaton contribution has the opposite sign to that
of the $\Box\phi (\partial_\mu \phi)^2$
term. Figure~\ref{fig3} shows the evolution of the solution for
$b=-c=-4$. The fixed point obtained by adding classical corrections to the
lowest-order action is well located as the evolution has already reached the  
$\dot{\bar{\phi}} \leq 0$ region of the $(H,\dot{\bar{\phi}})$ phase space,
indicating that generic loop corrections will drive
the evolution across the Einstein bounce. Indeed, setting $A=+4$ we see that the
evolution crosses the Einstein bounce as well as the $(-)$ branch. This suggests
that the violation of the NEC is too large and will not give an upper bound to
the curvature in the Einstein frame. An extra two-loop term is thus required to
instigate the transition to a decelerated expansion in the Einstein frame. As
shown in figure~\ref{fig3} when such a term is included with $B=-0.3$, we obtain 
a
successful implementation of the graceful exit. Once again, we invoke particle
production in order to eventually stabilise the dilaton in the usual FRW
cosmology.

An important issue concerns the sensitivity of our results to the values
of the parameters $a,b,c,d,A$ and $B$. Do we find successful transitions for
only a small range of these values, in which case we should be concerned that
our solutions are not representative of the typical case? Fortunately, we
have established that there does exist a large range of values of the parameters
which allow for successful exits between the two branches. Figure~\ref{fig0}
shows the range that lead to fixed point solutions when just the classical
corrections are included. From such a solution, it is then
relatively straightforward to achieve a
successful exit through the addition of the quantum corrections.

The role of $A$ and $B$ is less clear cut. Their actual values
determine the onset
when loop corrections become important, hence the value of $\phi$ when the
graceful exit is successfully completed. Generally we find that a successful
exit requires $A>0$ and $B<0$, implying that the two quantum loop terms
compete against each other in order to lead to a successful exit.
Unfortunately there is a degree of amBiguity present which arises because of
the invariance of the system of equations under a constant shift of $\phi \to
\phi + {\rm const}$, with a compensating shift of $A$ and $B$. The invariance
of the
system (up to a multiplicative constant in the action)
is manifest at the tree level and implies that we can make $\phi$
arbitarily
negative and guarantees a transition in the weak coupling regime. As mentioned
earlier, when the loop corrections are included, this simple invariance is
broken
and the shift in $\phi$ is compensated for by a shift in $A$ and $B$. For
example,
if we take the case $b=-c=-4$ and set $A=+4$ and $B=-0.3$,
we obtain a graceful exit with $\phi_{final} \simeq 0.75$. In shifting the 
dilaton by
$\phi_{*}=-1.5$, the equivalent dynamical evolution is obtained with a final 
value
$\phi_{final} \simeq - 0.65$ (weak coupling) with 
$A \to e^{2\phi_{*}} A=80.34$ and $B \to e^{4\phi_{*}} B=-121.03$.
There appears to be a price to pay, weak coupling seems to imply large loop
coefficicents. Normally we would expect
the coefficients of successive loops to be less important. 
This behaviour raises an interesting
question as to whether it is possible to have
genuinely weak coupling transitions with just loop corrections included. We have
tried this for a wide range of the fixed point solutions and found
the same type of behaviour.

\subsection{Entropic bounds}
There has recently been considerable interest in the possibility
that entropy considerations can provide new constraints on the allowed
evolution of the Universe \cite{Easther1,Veneziano9,Brustein3,Brustein4}.
Veneziano has suggested that in the context of string
cosmology, non-singular cosmologies should respect at all times a
Hubble entropy bound \cite{Veneziano9}. Brustein has proposed that such
solutions should satisfy a slightly different bound arising from a
generalised second law of thermodynamics (GSL) \cite{Brustein3}.
Do our non-singular solutions satisfy these bounds?

\begin{figure}[ht]
\begin{center}
\includegraphics[width=6.5cm]{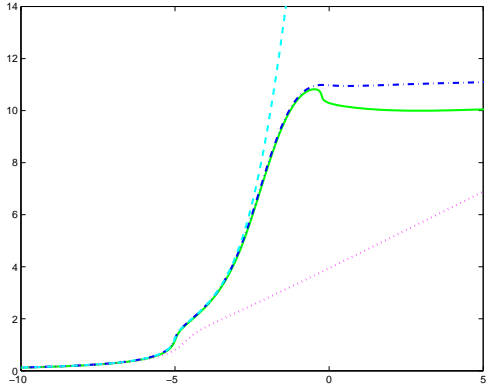}
\caption{$ln(S_{HB})_{t_{final}} - ln(S_{HB})_{t_{initial}}$ plotted as a 
function of $t_{final}$ for the setting $a=-1$, $b=-c=-4$ and $d=16$. The dotted 
line represents the classical correction, then the loop corrections with $A=+4$ 
(dashed line) and  $B=-0.3$ (dash-dotted line). The plain green line also 
includes the effect of particle creation, with $\Gamma_{\phi}=0.2$.}
\label{fig4}
\end{center}
\end{figure}

We will concentrate on
the example of the Hubble entropy bound \cite{Veneziano9}. Following
Bekenstein's work \cite{Bekenstein}, Veneziano suggested that for a
homogeneous cosmology, the radius of the largest black hole that can form is
determined by the largest causal scale available, namely the Hubble radius
$H^{-1}$. The maximum entropy enclosed in such a Universe corresponds to having
one black hole in a Hubble volume. Defining $n_H=a^{3} H^3$ to be the number of
cosmological horizons within a given comoving volume and $S_H=\vert H
\vert^{-2}e^{-2\phi}$ the maximal entropy within the horizon corresponding to a
black hole of radius $H^{-1}$, the Hubble entropy is given by $S_{HB} = a^{3} H
e^{-2\phi}$ \cite{Veneziano9}. From this it follows that:
\begin{equation}
\partial_{t}\;S_{HB} =   n_H \partial_t S_H + S_H \partial_t n_H. \label{H_ent}
\end{equation}
Enforcing that the rate of change of this geometric entropy $\partial_{t}
\;S_{HB} \geq 0$ leads to  the reduced inequality
\begin{equation}
\frac{\dot{H}}{H}- 2\dot{\bar{\phi}}\geq 0 \label{H_bound}
\end{equation}
with the lowest-order solutions of the PBB scenario saturating this geometric
bound \cite{Veneziano9,Bak,Kaloper4}.
Furthermore, it indicates that a fixed  point necessarily occurs for
non-positive $\dot{\bar{\phi}}$ (a conclusion that also follows from
the presence of a conserved quantity in the solutions \cite{Veneziano7}).
Brustein et al. provided evidence in \cite{Brustein4} that the bound was
satisfied for the non-singular solutions arising out of the purely classical
correction Eq.~(\ref{Class}) satisfying  $a=-1$ and the constraint
Eq.~(\ref{constraint}). We have confirmed this result, although we have also
found that all the non-singular solutions we have obtained, when including loop
corrections, lead to violations of this bound over short time intervals.
However, as pointed out by Veneziano, this is really a global bound, in the
asymptotic future, as we enter the FRW phase, we always find that $S_{HB}$
has increased. An example of this can be seen in figure~\ref{fig4}.

The comparison becomes a bit more difficult when considering the bound proposed
by Brustein arising from a
generalised second law of thermodynamics (GSL) \cite{Brustein3}. We do find a
class of solutions which satisfy such a bound for positive values of the
chemical potential $\mu$ that he introduced, but we also find solutions which
violate the bound. It is difficult to draw any real conclusion from this, not
least because we do not know the precise form of the quantum corrections or
particle production the true graceful exit solution will contain. It is
certainly interesting that there are regions of parameter space where the bound
is satisfied. The nature of these bounds is also under active consideration at
the moment \cite{VenBru}.

\section{Conclusions}
In this paper, we have obtained a class of non-singular cosmologies,
based on an effective action given
in Eqs.~(\ref{Low}), (\ref{Class}) and (\ref{efflag}).
The inclusion of such classical and
quantum corrections can lead to an evolution which smoothly joins
the inflationary pre-Big Bang solution with a decelerating FRW universe.
The classical correction is based on an enhanced form of the action which
includes up to four derivatives in the fields. The importance of quantum loops
in achieving a smooth  transition has become manifest, in agreement with
\cite{Brustein0,Brustein2}. Furthermore, we observe that these non-singular
solutions can satisfy the recently proposed entropics bounds when
loop-corrections are included.

Although encouraging, the solutions we have presented still have side effects;
in particular we had to stabilise the dilaton by hand. Also, it proved quite
difficult to obtain the transition in the weak coupling regime, whilst
keeping the loop corrections small. It is not clear to us, how serious
an issue this is as we do not know the form of the true corrections.  An
intriguing issue is the unusual behaviour surrounding the SFD case. Why do these
singular regions arise and what do they correspond to physically?

Finally, we would like to comment on a possible natural extension of this
work. Recently in refs
\cite{Durrer1,Durrer2} the authors have developed a
technique to determine the large-scale CMB anisotropy and power spectra
generated by massless axionic seeds in the pre-Big Bang scenario. They
numerically determined the CMB anisotropy power spectrum and pointed out the
differences ('isocurvature hump' at $l \sim 40$ and first acoustic peak at $l
\sim 300$) with more standard adiabatic models. These are
fascinating results, but are based on numerical solutions for the curvature
and dilaton that have not really avoided the curvature singularity. Instead they
are frozen at some scale, and then begin evolving again once the
post Big Bang FRW branch is entered. We are in a position to provide
solutions where the background fields evolve right through the transition
in a singularity free manner, and it would be useful to determine
how such an evolution impacts on the  modes leaving the horizon during the
transition period. How (if at all) do they influence the CMB spectrum at large
$l$? This is currently under investigation.

\acknowledgments{CC was supported by the Swiss NSF, grant No. 83EU-054774.
EJC was supported by PPARC. We are very grateful to Ramy Brustein and Gabriele
Veneziano for detailed discussions on the nature of the Entropy Bounds, to
Ruth Durrer and Malcolm Fairbairn for very useful comments, and to
the referee for detailed comments.}


\begin{thebibliography}{99}
\bibitem{Hawking1}S.W. Hawking and R. Penrose, Proc. Roy. Soc. Lond. {\bf A 314} 
(1979)
529.

\bibitem{ejc} J.E. Lidsey, D. Wands and E.J. Copeland, \hepth{9909061}.

\bibitem{Brandenberger1}R. Brandenberger and C. Vafa, \npb{316}{1989}{391}.

\bibitem{Veneziano1}G. Veneziano, \plb{265}{1991}{287}.

\bibitem{Tseytlin1}A. A. Tseytlin and C. Vafa, \npb{372}{1992}{443}.

\bibitem{Veneziano3}M. Gasperini and G.Veneziano, Astropart. Phys. {\bf 1} 
(1993) 317; \mpla{8} {1993}{3701}; \prd{50}{1994}{2519}.

\bibitem{Veneziano2} An updated collection of papers and references on the
pre-Big Bang scenario is available at "http://www.to.infn.it/teorici/gasperini/"

\bibitem{Veneziano4}A. Buonanno, T. Damour and G. Veneziano, 
\npb{543}{1999}{275}.

\bibitem{Gasperini1}M. Gasperini, \grqc{9902060}.

\bibitem{Weinberg} M. S. Turner and E. J. Weinberg, \prd{56}{1997}{4604}.

\bibitem{Kaloper} N. Kaloper, A. D. Linde, and  R. Bousso, 
\prd{59}{1999}{043508}.

\bibitem{Damour1}T. Damour and A.M. Polyakov, \npb{423}{1994}{532}.

\bibitem{Campbell1}B.A. Campbell and K.A. Olive, \plb{345}{1995}{429}.

\bibitem{Barreiro} T. Barreiro, B. de Carlos and E. J. Copeland, 
\prd{58}{1998}{083513}.

\bibitem{Brustein0}R. Brustein and R. Madden, \prd{57}{1998}{712}.

\bibitem{Veneziano5}R. Brustein and G. Veneziano, \plb{329}{1994}{429}.

\bibitem{Kaloper5}N. Kaloper, R. Madden and K. A. Olive, \npb{452}{1995}{677}.

\bibitem{Kaloper6}N. Kaloper, R. Madden and K. A. Olive, \plb{371}{1996}{34}.

\bibitem{Ellis}G.F.R. Ellis, D.C. Roberts, D. Solomons and P.K.S. Dunsby, 
\grqc{9912005}.

\bibitem{Veneziano7}M. Gasperini, M. Maggiore and G. Veneziano, 
\npb{494}{1997}{315}.

\bibitem{Foffa} S. Foffa, M. Maggiore and R. Sturani, \npb{552}{1999}{395}.

\bibitem{divers}I. Antoniadis, J. Rizos and K. Tamvakis, \npb{415}{1994}{497};
R.H. Brandenberger, R. Easther and J. Maia, \jhep{08}{1998}{007};
D.A. Easson and R.H. Brandenberger, \jhep{09}{1999}{003}.

\bibitem{Ghosh}A. Ghosh, R. Madden and G. Veneziano, \hepth{9908024}.

\bibitem{Brustein2}R. Brustein and R. Madden, \jhep{07}{1999}{006}.

\bibitem{Maggiore2}M. Maggiore and A. Riotto, \npb{548}{1999}{427}.

\bibitem{Meissner}K.A. Meissner, \plb{392}{1997}{298}.

\bibitem{Kaloper2}N. Kaloper and K.A. Meissner, \prd{56}{1997}{7940}.

\bibitem{Metsaev}R.R. Metsaev and A.A. Tseytlin, \npb{293}{1987}{385}.

\bibitem{Brustein1}R. Brustein and R. Madden, \plb{410}{1997}{110}.

\bibitem{Easther1}R. Easther and D. Lowe, \prl{82}{1999}{4967}.

\bibitem{Veneziano9}G. Veneziano, \plb{454}{1999}{22}.

\bibitem{Brustein3}R. Brustein, \grqc{9904061}.

\bibitem{Brustein4}R. Brustein, S. Foffa and R. Sturani, \hepth{9907032}.

\bibitem{Bekenstein}J.D. Bekenstein, \prd{23}{1981}{287}.

\bibitem{Bak}D. Bak and J.S. Rey, \hepth{9902173}.

\bibitem{Kaloper4}N. Kaloper and A. Linde, \prd{60}{1999}{103509}.

\bibitem{VenBru} R. Brustein and G. Veneziano, \hepth{9912055}.

\bibitem{Durrer1}R. Durrer, M. Gasperini, M. Sakellariadou and G. Veneziano,
\prd{59}{1999}{043511}.

\bibitem{Durrer2}A. Melchiorri, F. Vernizzi, R. Durrer and G. Veneziano, 
\prl{83}{1999}{4464}.

\end{thebibliography}
\end{document}